\documentstyle[aps,pre,epsfig]{revtex}

\begin{document}
\twocolumn

\title{Canonical and non-canonical equilibrium distribution}
\author{Mario Annunziato,$^{1,}$\footnote{annunzia@df.unipi.it}
 Paolo Grigolini,$^{1,2,3,}$\footnote{grigo@unt.edu}
 Bruce J. West,$^{4,}$\footnote{WestB@aro-emh1.army.mil}}
\address{$^{1}$Dipartimento di Fisica dell'Universit\`{a} di Pisa and
INFM,
Piazza Torricelli 2, 56127 Pisa, Italy}
\address{$^{2}$Istituto di Biofisica CNR, Area della Ricerca di Pisa,
Via Alfieri 1, San Cataldo 56010 Ghezzano-Pisa,  Italy}
\address{$^{3}$Center for Nonlinear Science, University of North Texas, \\
P.O. Box 305378, Denton, Texas 76203-5378}
\address{$^4$US Army Research Office, Research Triangle Park, NC}
\date{\today}
\maketitle

\begin{abstract}
We address the problem of the dynamical foundation of non-canonical
equilibrium. We consider, as a source of divergence from ordinary
statistical mechanics, the breakdown of the condition
of time scale separation between microscopic
and macroscopic dynamics. We show that this breakdown 
has the effect of producing a significant deviation from 
the  canonical prescription.
We also show that, while the canonical equilibrium can be
reached with no apparent dependence on dynamics, the specific form of
non-canonical equilibrium is, in fact, determined by dynamics.
We consider the special case where
the thermal reservoir driving the system of interest to equilibrium
is a generator of intermittent fluctuations. 
We assess the form of the non-canonical equilibrium reached by 
the system in this case. Using both theoretical and numerical 
arguments we demonstrate that L\'evy statistics are the best description 
of the dynamics and that the L\'evy distribution is the correct basin of 
attraction. 
We also show that the correct path to non-canonical equilibrium by means of
strictly thermodynamic arguments has not yet been found, and that further
research has to be done to establish a connection between 
dynamics and thermodynamics.
\end{abstract}

\pacs{05.45.+b,03.65.Sq,05.20.-y}

\section{Introduction}
\label{intro}

A growing interest in non-extensive thermodynamics can be traced back to the
1988 pioneering work of Tsallis \cite{tsallis1}. In particular, non-extensive
thermodynamics is the proper theoretical context for developing the
foundations of non-canonical distribution functions and non-canonical
equilibria in physical systems. On the basis of the subsequent development
of the original work by Tsallis, it has been argued that canonical
equilibria, the distributions that form the basis of equilibrium statistical
mechanics, are not generic. Rather, canonical equilibria are singular in a
more general form of equilibrium, called a generalized canonical equilibrium
\cite{tsallis2,brazil}. A number of recent papers elaborate on this 
idea \cite{walton,wilk,abe}, but we find
\cite{abe} to be of special interest addressing as it does the 
foundations of both
canonical and non-canonical equilibria. Rajagopal and Abe \cite{abe} 
prove that the
equilibrium described by the canonical distribution is not uniquely
determined by the micro-canonical distribution, as one finds in text books.
In fact if the phase space has a fractal, rather than smooth, structure, a
non-canonical distribution will result. However, although canonical
equilibria are not unique, and non-canonical equilibria are possible, one
might conclude on the basis of the arguments in \cite{abe} that the form of the
non-canonical distribution is uniquely of the form established by Tsallis
and his co-workers \cite{tsallis1,tsallis2}. 
As attractive as the probabilistic and entropic arguments of Ref. \cite{abe} 
are, they are not indisputable, and in fact we find herein, 
using dynamical arguments, that this is not the case.

  We show that the adoption of a dynamical
approach to thermodynamical equilibrium yields a different
conclusion. First of all, we argue
that the non-extensive condition based on memory, and probably
that resting on long-range correlations as well,
has the striking effect of making the role of dynamics
much more important than in the case of ordinary statistics.
The dynamic approach, in these non-extensive conditions,  generates
a form on non-canonical equilibrium that,  however, departs  from the
generalized canonical  form  prescribed  by  non-extensive 
thermodynamics. To make  it easier for the reader to follow our
arguments and to understand the purpose of the paper, illustrated in
Section \ref{purpose}, we shall first discuss the different natures  of the
entropic, dynamic and  stochastic  approaches  to
equilibrium.

\subsection{Non-extensive entropic indicator}
\label{non-ext}

The argument for the non-extensive thermodynamics of Tsallis goes as
follows. First of all, the conventional entropy of Gibbs
	\begin{equation}
	S(\Pi) \equiv - \int dx \; \Pi(x)\; ln \Pi(x) ,
	\label{normalentropy}
	\end{equation}
is replaced by the non-extensive entropic indicator
\begin{equation}
S_{q}(\Pi) \equiv  \int dx \;\frac { 1 -  \Pi(x)^{q}}{q-1}.
\label{nonextensiveentropy}
\end{equation}
Secondly, we have to apply a method of entropy maximization under given
physical constraints to determine the
most plausible shape of the unknown probability density function $\Pi(x)$
\cite{katz}.
   The first constraint is on the normalization of the distribution
   $\Pi(x)$:
   \begin{equation}
   \int dx \; \Pi(x) = 1 .
   \label{norm}
   \end{equation}
The second  constraint is on the first moment
of the variable $x$ itself.
   According to the most recent prescription of Ref.\cite{thirdrule}
   the constraints on $x$ must be applied\cite{note1} on the mean value
   $U_{q}$ defined by
	\begin{equation}
	U_{q} \equiv \frac{ \int dx \;x \;\Pi(x)^{q}}{\int dx \;\Pi(x)^{q} }.
	\label{firstmoment}
	\end{equation}
It has to be pointed out that in Ref.\cite{thirdrule} the physical
meaning of $x$ is that of energy. Here, we shall interpret $x$ as the
``coordinate'' of an overdamped particle driven by
a fluctuation-dissipation process
resulting from the interaction with a non-conventional ``thermal bath''.
When the friction term can be neglected we are expected to recover the
results of the earlier work of Ref.\cite{anna}.
Thus, for the same reasons \cite{paolonote}
as those illustrated in Ref.\cite{anna} we set a constraint on
the first moment of $|x|$.
The result of entropy maximization subject to the imposed 
constraints of Eq. (\ref{firstmoment}) yields
	\begin{equation}
	\Pi(x) = \left[1 - \frac{(1-q) \tilde{b} (x - U_{q})}{\int dx \;
	\Pi(x)^{q}}\right]^{1/(1-q)}/Z_{q},
	\label{generalizedcanonical}
	\end{equation}
where $Z_{q}$ is a normalization factor and $\beta$ is a constant value
stemming from the Lagrange multiplier associated with the constraint on
the variable $x$. Note that the expression of $\Pi(x)$ provided by
the authors of Ref.\cite{thirdrule} is not explicit. In fact, as
shown by Eq.(\ref{generalizedcanonical}), it is a functional 
of $\Pi(x)$. Thus, in
practice the explicit form of $\Pi(x)$ has to be established by means
of an iteration procedure. It is remarkable, however, that the
micro-canonical derivation from Ref.\cite{abe} results in the same
prescription as that of Eq.(\ref{generalizedcanonical}).
  From the point of view of the issues under discussion here
what matters is the fact that at the end of the iterative procedure
$U_{q}$ becomes a well-defined number. Consequently,
   the resulting expression for $\Pi(x)$ is a simple
analytical formula  that in the asymptotic limit, $|x| \rightarrow
\infty$, with the constraint on the first moment of $|x|$,
 has the same structure as that derived in
Ref.\cite{anna},
	\begin{equation}
	\Pi(x)={{\tilde{b}(2-q)}\over{[1+\tilde{b}(q-1)|x|]^{1/(q-1)}}}.
	\label{pi2}	
	\end{equation}
Note that the adoption of a constraint on the second moment would lead to
	\begin{equation}
	\Pi(x)=\frac{{\left[\frac{\tilde{b}(q-1)}{\pi}\right]}
	\frac{\Gamma(1/(q-1))} {\Gamma((3-q)/[2(q-1)])}}
	{[1+\tilde{b}(q-1) x^{2}]^{1/(q-1)}}.
	\label{new}
	\end{equation}

Note that both Eq.(\ref{pi2}) and Eq.(\ref{new}) are the result 
of an earlier prescription, missing the
normalization factor present in Eq.(\ref{firstmoment}). Nevertheless,
it has to be pointed out that both Eq.(\ref{pi2}) and
Eq.(\ref{generalizedcanonical}) share the characteristic
of having long tails with L\'{e}vy statistics.
In fact, it is well known \cite{MS84} that the anomalous
diffusion processes of the L\'evy kind, in the one-dimensional case,
are characterized by
probability distributions $p(x,t)$ whose Fourier transform in the
symmetric case reads
	\begin{equation}
	\hat{p}(k,t) = exp(-b|k|^{\alpha}t)  ,
	\label{pk}	
	\end{equation}
where $\alpha$ is the L\'evy index ranging, in principle, in the interval
$0 < \alpha \leq 2$, and $b$ denotes the diffusion intensity. The
inverse Fourier transform of $\hat{p}(k,t)$ of Eq.(\ref{pk}) is
characterized by the tail \cite{bruce}
	\begin{equation}
	\lim_{|x| \to \infty} p(x,t) \propto {t\over{|x|^{1+\alpha}}} \,,
	\label{lim}	
	\end{equation}
which would lead  immediately to the anomalous entropy index
	\begin{equation}
	q = 1 + 1/(1 + \alpha) .
	\label{anomalousentropicindex}
	\end{equation}
	\begin{figure}
	\begin{picture}(210,180)(0,0)
	\epsfxsize=6truecm
	\epsfxsize=8.5truecm
	\put(-10,0){\epsfbox{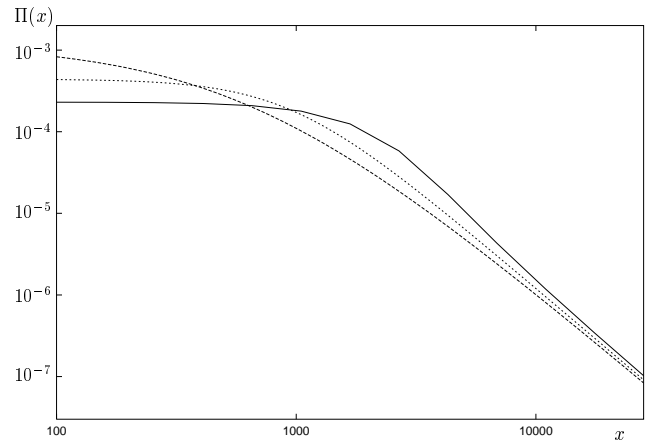}}
	\end{picture}
	\caption{Comparison of a L\'evy distribution (solid line)
		with distributions obtained by maximizing Tsallis 
		entropy (dashed lines). The L\'evy curve is obtained
		by the inverse Fourier transform L\'evy characteristic
		function with $\alpha=1.5$, $b=4.4\cdot 10^{4}$. 
		The asymptote is proportional to
		$1/x^{2.5}$. The long-dashed line comes from
		Eq. (\ref{pi2}) with the constraint on the first
		moment ($\tilde{b}=0.004$, $q=1.4$). The short-dashed
		line comes from Eq. (\ref{new}) with the
		constraint on the second moment ($\tilde{b}=1.4\cdot 10^{-6}$,
		$q=1.8$). The values of the parameter $\tilde{b}$ are 
		selected, in both cases, so as to fit the asymptotic behavior 
		of the L\'evy distribution. This constraint results in 
		significant departures from the L\'evy distribution at small 
		and intermediate distances.}
	\label{ConfrontoLevyTsallis}
	\end{figure}
We have to stress, however, that except for the case $\alpha = 1$,
corresponding to the ballistic limit, the central part of the
distribution produced by the L\'{e}vy process of Eq.(\ref{pk})
cannot be expressed in an analytical form and can
   significantly depart from the analytical form of both Eqs.(\ref{pi2})
   and (\ref{generalizedcanonical}). The adoption of the prescription
   of Eq.(\ref{new}) is expected to yield better agreement, but, as we
   shall see subsequently, it does not seem to be compatible with
   the nature of the dynamical approach to equilibrium.
   In Fig.\ref{ConfrontoLevyTsallis} we compare the 
   L\'{e}vy distribution to both the prediction
   of Eq.(\ref{pi2}) and Eq.(\ref{new}) and find that there exists a
   significant disagreement between Tsallis and L\'{e}vy statistics,
   even though some authors\cite{wilk} refer to them as equivalent.
In this paper we focus our attention on the origin of the difference
between the two kinds of statistics.
Our dynamic approach leads to a form of equilibrium that is stable, in 
the sense of the L\'evy-Gnedenko theorem \cite{gnekol}, while the generalized 
canonical equilibrium of Eq.(\ref{generalizedcanonical})  is not.  
This is evident in the free case, 
due to the difference between the
Tsallis and the L\'{e}vy structure. In Section \ref{theo} we prove 
that this is so also in the presence of a feedback with the system of
interest of the generator of fluctuations.

\subsection{From dynamics to thermodynamics}
\label{dyntotherm}
We are convinced that there are no
incontrovertible reasons why the canonical distribution should be the
unique form of thermodynamic equilibrium, and on this issue
we essentially agree with the point of view of Rajagopal and
Abe\cite{abe}. 
However, we are equally convinced that the generalized canonical form
of Eq.(\ref{generalizedcanonical}) does not satisfy the stable conditions
necessary for L\'{e}vy statistics, and the results of the
present paper can be thought of as providing plausible evidence of that.
To substantiate this view with dynamical arguments, it
is convenient to concisely review the results of an earlier
work\cite{bianucci}.

The ambitious purpose of this earlier work was that of reversing
the path from thermodynamics to mechanics established by Boltzmann.
The main idea behind Ref.\cite{bianucci} is as follows. The
Fokker-Planck equation is a well-known 
description of the evolution of the probability density in
phase space.
Many attempts have been made in the literature to derive
this equation\cite{ford,ullersma,philipson,vitali,lindenberg}.
However,
all these attempts rest on the assumption that the bath, responsible for
the Brownian behavior of the particle of interest, is given by a set
of harmonic oscillators. 
This means that a statistical assumption must
be made on the initial condition of the bath, which is arbitrarily
given a canonical equilibrium distribution, corresponding to a given
temperature $T$ of the thermal bath.

Consequently, this kind of approach to Brownian
motion is only partially dynamical, since significant use is already
made of statistical mechanics, and thus of thermodynamics
\cite{lindenberg2}.
The authors of Ref.\cite{bianucci} adopted a totally different approach.
They assumed that a given oscillator, playing the role of a stochastic
Brownian
particle, interacts with another Hamiltonian system, which should play
the role of a bath. 
They called this second system a \emph{booster}, to stress that
the ensuing approach has to rest only on the dynamical properties of
this kind of bath, with no use whatsoever of thermodynamical arguments.
After establishing the dynamical conditions ensuring the validity of
the Fokker-Planck equation, it is found that the oscillator
of interest reaches a canonical equilibrium distribution. 
Due to the
nature of the procedure adopted the width of this canonical
distribution  depends only on the parameters of the Hamiltonian system
under study. Consequently, it is possible to derive a mechanical
expression for temperature. This is the key result of
Ref.\cite{bianucci}, which reads
	\begin{equation}
	k_{B}T = \left[\frac{\partial}{\partial E} ln A(E)
	+ \frac{ \partial}{\partial E} ln
	\{\langle \xi^{2}\rangle_{eq} 
	Re[\hat{\Phi}_{\xi}(\omega)]\} \right]^{-1}.
	\label{bianucci}
	\end{equation}
Note that $\xi$ denotes the \emph{doorway} variable, namely the
variable of the booster through which the interaction between the booster
and the oscillator of interest is established.
The symbol $\hat{\Phi}_{\xi}(\omega)$
denotes the
Laplace transform of the correlation function of $\xi$ evaluated at
the oscillation frequency of the oscillator. 
The structure of this
expression reflects the application of linear response
theory\cite{bianucci}.
The correlation function whose Laplace transform is in Eq. (\ref{bianucci})
is evaluated assuming the booster to be in a micro-canonical
equilibrium with energy $E$ and this condition is not affected by the
interaction with the oscillator.  
The symbol $A(E)$ denotes
the number of states of the booster in the same physical condition,
and, consequently, obeys the ordinary prescription \cite{huang}
	\begin{equation}
	A(E) \propto E^{N/2},
	\label{huang}
	\end{equation}
where $N$ is the number of degrees of freedom in the booster.

The authors of Ref.\cite{bianucci} note that for $N \rightarrow \infty$,
	\begin{equation}
	\frac{ \partial}{\partial E} ln \{\langle \xi^{2}\rangle_{eq}
	Re[\Phi_{\xi}(\omega)]\}
	<< \frac{\partial}{\partial E} ln A(E) .
	\label{crucialinequality}
	\end{equation}
This means that in the limiting case of infinitely many degrees of
freedom the predictions of Boltzmann are recovered:
$1/(k_B T) = \frac{\partial}{\partial E} ln A(E)$.
Note that the condition of Eq.(\ref{crucialinequality}) holds true if
the doorway variable $\xi$ depends on a number of particles which is
kept fixed with $N \rightarrow \infty$. In other words, the condition
of Eq.(\ref{crucialinequality}), or equivalently,
Boltzmann's principle, rests on an interaction condition,
which is crucial for the extensive statistical mechanics perspective
to apply. 

Let us see this aspect in detail. In Ref.\cite{bianucci}
the booster is the the well known Fermi-Pasta-Ulam (FPU) system\cite{fermi}.
This is a  chain of particles
interacting with one another via nonlinear interactions. Let us
consider two opposite conditions. In the former, which is the one adopted
in
Ref.\cite{bianucci}, the oscillator of interest interacts only with
the first particle of the FPU chain. 
In the latter the oscillator of
interest interacts with all the particles of the FPU chain, with
interaction strength of comparable intensities. The former condition
refers to a short-range interaction, confined to the positions of the
oscillator of interest and to the first particle of the FPU chain. In
the latter condition, the interaction extends over the whole FPU chain.
It is evident that the former condition fits the inequality of
Eq.(\ref{crucialinequality}), whereas the latter does not.
This means that the latter condition results in a non-extensive form
of dynamics, with a consequent breakdown of the prescriptions of
ordinary statistical mechanics. The former condition, on the contrary,
for $N \rightarrow \infty$, recovers ordinary statistical mechanics.
This suggests that the dynamical corrections to the Boltzmann
principle, recorded in Ref.\cite{bianucci} for relatively small values
of $N$, are a manifestation of  incipient non-extensive statistical
mechanics, and so are very close to the breakdown of the Fokker-Planck
treatment on which the analysis Ref.\cite{bianucci} rests.
Note that when we assign to the oscillator of interest a very low
frequency, compared to the booster frequencies, the
quantity $Re[\hat{\Phi}_{\xi}(\omega)]$ turns out to virtually coincide 
with the time scale
$T_{micro}$ of the variable $\xi$, defined by
	\begin{equation}
	T_{micro} \equiv \int_{0}^{\infty}\Phi_{\xi}(t)\;dt < \infty  .
	\label{finitetime}
	\end{equation}
Herein we focus our attention on the case where
	\begin{equation}
	lim_{t \rightarrow \infty} \Phi_{\xi}(t) = const/t^{\beta},
	\label{timebreakdownofextensivity}
	\end{equation}
with
	\begin{equation}
	0 < \beta < 1  .
	\label{crucialpowerinterval}
	\end{equation}
The inverse power law form of the correlation function
means that we select time memory as the source of violation of extensivity, 
rather than long-range spatial interactions. 
The parameter $T_{micro}$ denotes the correlation time of
the fluctuating variable $\xi$. Here we consider the dichotomous case,
where the variable $\xi$ has two distinct values, $W$,  and $-W$, with
fluctuating time durations. The correlation function $\Phi_{\xi}(t)$ is
proven \cite{allegro} to be proportional to the second time derivative of the
distribution of waiting times in the two states of the variable $\xi$.
This function, as we shall see in Section \ref{stocdyn},
depends on another time, $\bar{T}$, as well as on $\beta$.

In conclusion, the work of Ref.\cite{bianucci} establishes the dynamical
conditions necessary to derive canonical equilibrium. The authors prove that
canonical equilibrium implies an interaction with a booster with a
finite time scale and a short-range interaction. We can also
observe that in this case the use of
dynamical arguments yields the same conclusions as the very simple
argument based on the law of large numbers. 
The advocates
of the law of large numbers for the foundation of statistical
mechanics\cite{lebowitz} might judge the dynamical perspective
to be of limited use. We prefer to interpret the conclusion of 
Ref.\cite{bianucci}
as proof that in the extensive case the canonical 
distribution description of thermodynamic equilibrium is
unique. The numerical work of Ref.\cite{bianucci} shows that
it is very difficult to detect the dynamical corrections to the
Boltzmann principle illustrated by Eq.(\ref{bianucci}) from within
an extensive perspective. This is so because the dynamical
corrections become significant when the booster is small, and, as
clearly pointed out by the recent work of
Gross\cite{gross}, a small system is
essentially non-extensive. Thus, the numerical findings of
Ref.\cite{bianucci} refer to a condition very close to
the breakdown of the extensive condition and consequently  of
the canonical equilibrium on which Eq.(\ref{bianucci}) rests.
This means that the dynamical approach can be of great utility. 
In the case of non-extensive statistical mechanics,
the dynamic approach is probably the only
non-ambiguous way to address  the subtle issues
posed by the entropic and probabilistic methods.

   \subsection{Stochastic dynamics}
   \label{stocdyn}
   In this section we explain the nature of our dynamical approach to
   the non-canonical equilibrium distribution. This approach rests
   on a stochastic method adapted to the need of
   realizing a dichotomous variable 
   $\xi$ with the two possible values,
   $W$ and $-W$, and with the waiting time
   distribution
   	\begin{equation}
	\psi(t) = \frac{(\beta \bar{T})^{\beta+1}(\beta + 1)}
		{(\beta \bar{T} + t)^{2 + \beta}},
	\label{waitingtimedistribution}
	\end{equation}
   where $\beta$ ranges in the interval of
   Eq.(\ref{crucialpowerinterval}). Note that the parameter $\bar{T}$ is of
   crucial importance to define the time scale of our process
   and corresponds to the mean residence time in either of the two
   states of the velocity variable $\xi$.
   In accordance with the prescriptions of Ref.\cite{allegro} this
   waiting-time distribution yields the kind of correlation function
   $\Phi_{\xi}(t)$ that we plan to study herein
   (see Eq.(\ref{timebreakdownofextensivity})). In fact, as shown in
   Ref.\cite{allegro}, the form of Eq.(\ref{waitingtimedistribution})
   yields the correlation function
   	\begin{equation}
	\Phi_{\xi}(t) = \frac{(\beta \bar{T})^{\beta}}
	{(\beta \bar{T} + t)^{\beta}},
	\label{completeform}
	\end{equation}
   fitting the asymptotic time limit of Eq.(\ref{timebreakdownofextensivity}).
   This condition can be considered as the natural one-dimensional
   counterpart of the two-dimensional billiards of
   Zaslavsky \cite{zaslavsky}. 
   From this point of view, our commitment
   to the adoption of a merely dynamical approach is not broken, since
   the stochastic approach that will be adopted in Sections \ref{dynamical}
   and \ref{theo} is
   statistically equivalent to the adoption of the intermittent map
   of Ref.\cite{geisel}, on which the theoretical work of
   Ref.\cite{anna} is based.

   We stress that the equivalence between a dynamical map and a 
   stochastic process is the reason why contact can be
   established between the dynamical and the entropic approaches. In fact,
   as shown in Ref.\cite{anna} the non-extensive Tsallis entropic indicator
   serves the purpose of guessing the most convenient form
   for the transition probability $\Pi(x)$, which is then related
   to the waiting-time distribution
   $\psi(t)$ through the basic property
	\begin{equation}
	\label{pipsi}
	\Pi(x) = \frac{1}{2}\psi(x/W)/W .
	\end{equation}
The factor of $1/2$ takes into account that the probability of making the
jump $x$ is equal to that of making the jump $-x$, namely, the jump in
the opposite direction.
The authors of Ref.\cite{anna} show that the left-hand term of this
equality, Eq.(\ref{pipsi}), can be predicted using entropic 
arguments, while the
right-hand term of the same equation is dictated by dynamical arguments
based on the intermittent map of Ref.\cite{geisel}.
These dynamical arguments are supplemented by
the assumption of random injection of the trajectory from the
chaotic into the laminar region of the intermittent map, an argument 
leading to an analytical
prediction for $\psi(t)$ in complete agreement with the numerical
observation of dynamics\cite{geisel}.

The earlier work of Ref.\cite{anna} established that the
adoption of the method of entropy maximization applied to the
non-extensive entropy of Eq.(\ref{nonextensiveentropy}) results in a
form of $\Pi(x)$ which is compatible with the birth of L\'{e}vy
statistics. 
However, the $\Pi(x)$ thus derived is not the equilibrium
distribution of the variable $x$. Rather it is the probability for the
random walker to make a jump of length $|x|$. This is a stationary
property determined by the special kind of booster here under study.
The ensuing diffusion process yields a L\'{e}vy form as a result of
the L\'{e}vy-Gnedenko theorem \cite{gnekol}. 
As shown in Fig. \ref{ConfrontoLevyTsallis}, the shape of this distribution
departs from the form of the generalized canonical distribution of
Eq.(\ref{generalizedcanonical}), even if we adopt the constraint on
the second moment, 
in spite of the fact that it does not fit the nature
of the dynamical approach here illustrated.

To account for this discrepancy we might make the conjecture
that the comparison between the L\'{e}vy statistics and the Tsallis
generalized canonical distribution is not appropriate. The former refers to a
diffusion process, and the latter to an allegedly equilibrium
condition. 
Actually, after exploring this possibility we shall conclude that the 
latter, at least in the case of the dynamic model of the present paper, 
cannot reflect an equilibrium property.
However, at the present stage, we are forced
to develop a picture comparable to that of the earlier work of
Ref.\cite{bianucci}. We have to study a case where the variable $x$
not only undergoes the influence of the diffusion producing fluctuations,
but it produces a feedback on its own ``bath'',
balancing the diffusion process, so as to create an equilibrium condition. 
This is the
condition to compare to the generalized canonical distribution of
Eq.(\ref{generalizedcanonical}). In other words, to address, from a
dynamical perspective, the issues recently dealt with by Rajagopal and
Abe\cite{abe} we cannot disregard the feedback from the system to the
``bath.''

\subsection{Purpose of the present paper}
\label{purpose}
At this stage it is much easier for us to illustrate the main purpose
of the present paper. We want to explore a condition where the
booster does not fulfill the key condition of Eq.(\ref{finitetime}),
that is, the microscopic relaxation time diverges,
and we want to assess whether or not this divergence leads to 
the generalized canonical distribution of Eq.(\ref{generalizedcanonical}).
We aim at answering the question: Which is the form of the
equilibrium distribution reached by the system of interest when the
time scale of the \emph{booster} is infinite?

To answer this question we do not adopt, as done in Ref.
\cite{bianucci}, a Hamiltonian approach. The latter approach is difficult
for obvious reasons. A numerical simulation check of the theoretical 
prediction would imply technical difficulties caused by the slowness 
of the booster itself. This means that the anomalous booster is replaced, as
done in earlier papers\cite{juri,mario}, by a generator
of a dichotomous fluctuation with an inverse power law distribution of
waiting times. The essential ingredient of the approach of
Ref.\cite{bianucci}, not yet present in the dynamical derivation of
a free L\'{e}vy diffusion, is a feedback of the diffusing variable on
the generator of the fluctuations. This aspect has already been
considered in the earlier work of Ref.\cite{juri}. 
However, in that paper
the dynamical approach to the feedback was replaced by a
phenomenological friction, which did not allow the authors of that
paper to keep the promise of resting solely on dynamical
arguments at any level.
In conclusion, we adopt the program of Ref.\cite{bianucci}, based on
the observation of the fluctuation-dissipation process
caused by the interaction of a particle with a booster 
having no finite time
scale. The presence of feedback serves the purpose of balancing
the diffusion process with dissipation so as to result eventually 
in an equilibrium condition.

With the program of Ref.\cite{bianucci} in mind, we have to refer
ourselves to the correlation function, Eq.(\ref{completeform}),
with the index $\beta$ fulfilling the
condition of Eq.(\ref{crucialpowerinterval}), and so implying the breakdown
of the time scale separation between the macroscopic and the
microscopic levels. 
The realization of the program of
statistical mechanics requires an accurate definition of the
process of memory erasure associated with the transition from one to
the other velocity state\cite{mauro}. This is more conveniently
defined by the waiting-time distribution $\psi(t)$ of
Eq.(\ref{waitingtimedistribution}), than by
the correlation function of Eq.(\ref{completeform}). We see that even
if the condition of Eq.(\ref{crucialpowerinterval}) applies, the time
$\bar{T} = \int_0^\infty t\; \psi(t)\, dt$ remains finite. 
Memory of microscopic dynamics is lost in times $t>>\bar{T}$.

At this stage it is convenient to support our claim concerning memory
erasure with arguments borrowed from the earlier investigation of
Gaspard and Wang\cite{gaspard}. These authors studied the Kolmogorov
complexity of the Manneville map and found that in the regime
corresponding to the dynamical foundation of L\'{e}vy processes the
Kolmogorov complexity is a linear function of time. This  means that
at a given time $t$ the number of transitions $M$ from the one to the
other laminar region is given by $M \propto t/\bar{T}$. 
This means that
for $t \rightarrow \infty$ the conditions for the realization of the
generalized version of the central limit theorem are fulfilled, since
the function defined by
	\begin{equation}
	p(x,M) \equiv \Pi(x)*\Pi(x)*\ldots*\Pi(x)*p_{in}(x),
	\label{generic}
	\end{equation}
where the asterix denotes a convolution, and for $M\rightarrow \infty$, 
Eq. (\ref{generic}) tends to the L\'{e}vy distribution.
One might be tempted to make the conjecture that in the presence of
feedback the number $M$ cannot increase beyond some limit, and that
equilibrium is reached with a relatively low value of $M$ so
as to allow the distribution to maintain the structure of a
generalized canonical distribution \emph{\'{a} la }Tsallis. 
In this paper we limit our analysis to the case
where the linear response theory of Section \ref{theo} holds true. This forces
us to adopt a feedback so weak that this possibility does not emerge
from our simulations.

\section{Dynamical model}
\label{dynamical}
The dynamical model studied here depends on a process of free diffusion
with feedback, established through control of a dynamical
parameter of the booster. This feedback process has the role of
balancing diffusion so as to realize dynamically the equilibrium
condition. We illustrate first the dynamical model used as the generator
of free diffusion, then we show how the feedback is realized, and
finally we illustrate the numerical technique adopted.

\subsection{Free diffusion}
\label{free}
Formally free diffusion is realized using the equation of motion
	\begin{equation}
	\frac{dx(t)}{dt} = \xi(t)   .
	\label{fundamentalequation}
	\end{equation}
Here the dynamical variable $x(t)$ denotes either a spatial coordinate
or a velocity. In the former case the variable $\xi(t)$ has to be
considered a fluctuating velocity, while in the latter case
it is regarded as a fluctuating acceleration. The results are
equivalent, and the reader can adopt either of them, even if we 
consider $x(t)$ to be a spatial coordinate so as to make
the connection with earlier work \cite{anna} more natural.

We assume the variable $\xi(t)$ to be dichotomous, namely, we shall
assign to this variable only two distinct values, either $W$ or $-W$.
The motivation for this choice is not only simplicity,
but also has to do with the main purpose of this paper, that of
assessing the specific form the non-canonical equilibrium. Also,
of course, and prior to this purpose, that of proving that
deviations from canonical equilibrium
can be generated as a consequence of the breakdown of the ordinary
condition of time-scale separation between macroscopic and
microscopic processes. In the limiting case where the variable
$\xi(t)$ is Gaussian, and consequently produces a Gaussian equilibrium
distribution \cite{grigo}, the resulting diffusion process is Gaussian
and the final equilibrium distribution 
can turn out to be Gaussian
as well \cite{juri}, as a result of having established a seed of 
ordinary statistics at the microscopic level.
For this reason the choice of 
microscopic statistics, departing from the ordinary statistical
condition is crucial, and the adoption of the dichotomous assumption
serves the purpose of eventually establishing a non-canonical
equilibrium.

 The variable $\xi(t)$ keeps one of the two possible values for times
with a random duration. Thus, a statistical treatment must be adopted
and a waiting-time distribution $\psi(t)$ is used. The time
intervals of sojourn in a given state are labelled by an integer
index $k$ running from $1$ to $+ \infty$. At the end of
any sojourn the variable can either change its value, and thus
make a transition from $W$ to $-W$ or from $-W$ to $W$, or it can
keep its original value. The probability of changing values and that of
keeping the same value are equal, and thus equal to $1/2$.  We can
interpret the variable $\xi(t)$ as a function of the continuous time
$t$ by adopting the following prescription:
	\begin{equation}
	\xi(t) = \sum_{k} \xi_{k} \chi_{t_{k}, t_{k+1}}(t),
	\label{discretecontinuous}
	\end{equation}
where the times $t_{k}$ and $t_{k+1}$ denote the beginning and the
ending time of the $k-th$ sojourn, respectively, and the values
$\xi_{k}$ are randomly assigned either the value $W$ or the value $-W$
with equal probability and $\chi_{t_k,t_{k+1}}(t) =1$ 
if  $t_k \leq t < t_{k+1}$.
The sequel of the times $t_{k}'$ is fixed by the waiting time density
distribution $\psi$, by selecting a given value $\tau_{k}$, of the
interval $[\tau_{k}, \tau_{k} + \epsilon]$, with $\epsilon << 1$ with
the probability $\psi(\tau_{k}) \epsilon$. 
The waiting time
distribution $\psi$ is assigned an inverse power law form determined
by the constraint of yielding the correlation function
	\begin{equation}
	\Phi_{\xi}(t) = \frac{(\beta \bar{T})^{\beta}}
		{(\beta \bar{T} + t)^{\beta}},
	\label{correlationfunction}
	\end{equation}
where $\beta$ is a positive number determining the
integrability of the correlation function. If $\beta >1$  the
correlation function, Eq.(\ref{correlationfunction}), is integrable
and the microscopic time scale
	\begin{equation}
	T_{micro}  \equiv  \int_{0}^{\infty} \Phi_{\xi}(t) dt
		= \frac{\beta \bar{T}}{\beta - 1}
	\label{microscopictime}
	\end{equation}
can be defined.
According to Refs.\cite{thechoiceofmario,allegro} the form of
$\psi(t)$ is determined by that of $\Phi_{\xi}(t)$ through the relation
	\begin{equation}
	\psi(t) = \bar{T} \frac{d^{2}}{dt^{2}} \Phi_{\xi}(t) ,
	\label{fromwaitingtocorrelation}
	\end{equation}
which yields
	\begin{equation}
	\psi(t) = \frac{(\beta + 1)(\beta \bar{T})^{\beta + 1}}
		{(\beta \bar{T} + t)^{\beta + 2}}.
	\label{explicit}
	\end{equation}
We note that the parameter $\bar{T}$ appearing in both
Eq.(\ref{correlationfunction}) and Eq.(\ref{explicit}) is the mean
sojourn time. Thus, in some sense, when the microscopic time of
Eq.(\ref{microscopictime}) becomes infinite, the role of the microscopic
time scale is played by $\bar{T}$.

By integration of Eq.(\ref{fundamentalequation}) we get
	\begin{equation}
	x(t) = \xi_{n}(t - t_{n-1}) + \sum_{k = 0}^{n-1} \xi_k \tau_{k},
	\label{xattimet},
	\end{equation}
where $t$ is a time located in the interval $[t_{n-1}, t_{n-1}
+\tau_{n}]$ with $t_{n-1} = \sum_{k=0}^{n-1} \tau_{k}$, in accordance
with the earlier prescriptions. 
Furthermore, the transition probability of
Section \ref{non-ext} is obviously related to the waiting time 
distribution by Eq.(\ref{pipsi}). 
Note that, as discussed in detail in Refs.\cite{juri,mario,tesimario}, 
there are two distinct basins of attraction for the diffusion process
resulting from the repeated occurrence of the transition process of
Eq.(\ref{pipsi}), the Gauss basin for $\beta > 1$ and
the L\'{e}vy basin for $\beta <1$.

\subsection{The booster linear response and the diffusion control}
\label{linear&diffusion}
To control diffusion in such a way so as to generate a known equilibrium
distribution we adopt a model based on replacing
Eq.(\ref{fundamentalequation}) with
	\begin{equation}
	\frac{dx(t)}{dt} = \xi_{x}(t),
	\label{withfeedback}
	\end{equation}
which means that the motion of the fluctuating variable $\xi_{x}(t)$ 
also depends on the value of the dynamic variable $x$. We make the 
simplifying assumption that the \emph{relaxation time}, or response 
time, of the booster, $T_{B}$, is shorter than that of the  
particle of interest, $T_{R}$, so as to ensure for the booster a 
\emph{time dependent equilibrium condition} determined by the state of 
the system of interest. In other words, the booster is assumed to be in 
a condition of equilibrium determined by the variable $x$. As the 
variable $x$ moves from an initial condition $x(0)$ out of equilibrium state,
i.e. from a position larger than that fluctuations of $x$ itself, to the 
final equilibrium, the booster correspondingly moves from an 
equilibrium to another equilibrium condition. All this is illustrated 
by the following mathematical arguments. These arguments do not aim at 
providing a rigorous mathematical treatment, but rather a heuristic 
treatment reflecting the physical condition that we are assigning to 
the booster.

To make the analysis in terms of a strictly stochastic process,
we  build up the following regularized process
$\tilde{x}_{T_\rho}(t)$, defined by:
%
%
	\begin{eqnarray}
		\frac{ d \tilde{x}_{T_\rho}(t)}{dt} & \equiv & 
		\frac{x(t) - x(t-T_\rho)} {{T_\rho}} = 
             	\frac{1}{{T_\rho}} \int^t_{t-{T_\rho}} \xi_x(t')\, dt' \equiv
		\nonumber \\
		& \equiv & \frac{1}{{T_\rho}} \int^t_{t-{T_\rho}} 
		\xi_{\tilde{x}}(t')\,dt'\, ,
	\label{Processo_Regolarizzato}
	\end{eqnarray}
where ${T_\rho}$ is a regularization time and in the last
equivalence it has been explicited the spatial dependence of 
the noise on $\tilde{x}$. This new
process $\tilde{x}$ has the characteristics of $x(t)$ in the mean, but its
fluctuations are smoothed by using the temporal mean.

If the response time $T_B$ of the bath is fast and if ${T_\rho}$ 
is long enough, but always smaller than relaxation time $T_R$
of the variable  $x$, one can suppose that the phase space mean is
almost equivalent to the temporal mean:
%
	\begin{equation}
		\frac{d \tilde{x}_{T_\rho}(t)}{dt}= 
        	\frac{1}{{T_\rho}} \int^t_{t-{T_\rho}} \xi_{\tilde{x}}(t') dt' 
		\sim \langle \xi_{\tilde{x}} \rangle.
	\label{MediaTemporale&Spaziale}
	\end{equation}
Now, assuming linear response is valid, 
the mean value of the thermal bath variable depends on $\tilde{x}$ as:
%
%
	\begin{equation}
		\langle \xi_{\tilde{x}} \rangle = - \gamma \tilde{x}.
		\label{IpotesiRispostaLineare}
	\end{equation}
With the linear response hypothesis inserted into 
Eq. (\ref{MediaTemporale&Spaziale})
the equation for the regularized process
%
%
	\begin{equation}
		 \frac{d \tilde{x}_{T_\rho}(t)}{dt} \sim
		 -\gamma \tilde{x}_{T_\rho}(t)
	\label{Smorzamento}
	\end{equation}
is obtained. The solution $\tilde{x}_{T_\rho}(t)=x(0) e^{-\gamma t}$
shows a relaxation toward the equilibrium  $x \simeq 0$, on 
a time scale $T_R \simeq 1/\gamma$.

The analogy with a stochastic equation, such as the Ornstein-Uhlenbeck 
or its extension to L\'evy noise \cite{seshadri}, is completed by adding 
a fictitious stochastic noise $d \tilde{\xi}_{T_\rho}(t)$ in 
(\ref{Smorzamento}), that summarizes 
the fluctuations on the time scale ${T_\rho}$ and that 
compensates for the approximation inserted into Eq. 
(\ref{MediaTemporale&Spaziale}):
%
%
	\begin{equation}
		 d \tilde{x}_{T_\rho}(t) = 
		-\gamma \tilde{x}_{T_\rho}(t) dt + d \tilde{\xi}_{T_\rho}(t)
	\label{Analogia}
	\end{equation}
Hence, with the assumptions of linear response theory and relaxation times
of the thermal bath faster than of that of the variable of interest
(partial equilibrium), a diffusive process becomes an equilibrium
process.
We emphasize that the analogy  (\ref{Analogia}) is true only if:
	\begin{equation}
	T_{B}<< T_{\rho}<< T_{R}=\gamma^{-1}.
	\label{inequalities}
	\end{equation}
By using this simplified analysis it is also possible to make
qualitative predictions about the equilibrium distribution.

When $x \simeq 0$, the feedback $-\gamma \tilde{x}_{T_\rho}(t)$
can be neglected with respect to the fluctuations 
$\tilde{\xi}_{T_\rho}(t)$. This limit corresponds to free diffusion 
for which the central limit theorem (generalized or not) can be used.
The distribution is Gaussian or L\'evy's according to whether the
correlation function is integrable or not.

When $|x| >> 0$ the feedback term (\ref{IpotesiRispostaLineare})
prevails over fluctuations in Eq.(\ref{Analogia}). Then the 
dynamical variable is pushed 
back to values near the equilibrium  $x \sim 0$ as an attractive
field and, so, quenching the diffusion. 
This description corresponds, at least for the region $|x|<<W/\gamma$,
to what one could obtain if in Eq. (\ref{Analogia}) the 
fictitious noise $\tilde{\xi}_{T_\rho}(t)$ is simply stochastic
\cite{seshadri}.

The condition (\ref{IpotesiRispostaLineare}) is satisfied when 
the time evolution of $\xi_{x}$ is replaced by  the
stochastic variable $\xi_{k}$ appearing
in Eq. (\ref{discretecontinuous}) with the
the stochastic variable  $\xi_{k}(x_{k})$, which has two distinct
values, with the same probability:
%
%
	\begin{equation}
	\xi_k(x_k) \equiv 
	\left\{
		\begin{array}{c}
		W_+(x_k) =   W - \gamma x_k \\
		W_-(x_k) = - W - \gamma x_k 
		\end{array}
	\right.
	\label{StatiDicotomiciConAttrito}
	\end{equation}
In the time scale $T_{\rho}$, we can replace the
variable $\xi_{x}(t)$ of Eq.(\ref{withfeedback}) with its average
over the faster bath fluctuations, and thus with
	\begin{equation}
		\langle \xi_{x}\rangle = \frac{1}{2} 
		[W - \gamma x - (W + \gamma x)] = -\gamma x .
		\label{regularization}
	\end{equation}

	\begin{figure}
	\begin{picture}(210,180)(0,0)
	\epsfysize=6truecm
	\epsfxsize=8.5truecm
	\put(-10,10){\epsfbox{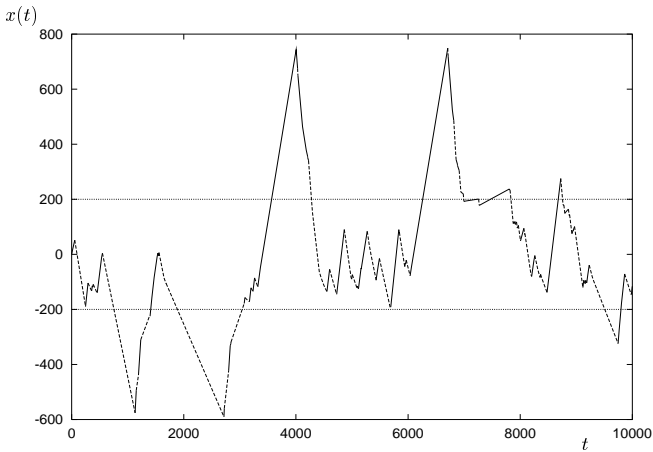}}
	\end{picture}
	\caption{Sample trajectory for the diffusion process with feedback
		defined in Eq. (\ref{withfeedback}). The
		solid and dashed lines denote the states $W_{+}$ and $W_{-}$,
		respectively. See Eq. (\ref{StatiDicotomiciConAttrito})
		for the definition of these two states. The
		two horizontal lines defining the central stripe 
		correspond to the
		levels $W/\gamma$ and $-W/\gamma$. The parameters used 
		have the values:
		$\gamma = 5\cdot 10^{-3}$, $W = 1$, $\beta = 0.5$, 
		$\bar{T} = 50$.
		}	
	\label{CampioneProcessoConRetroazione}
	\end{figure}
In Fig. \ref{CampioneProcessoConRetroazione} we show a sample 
trajectory corresponding to the dynamic prescriptions of 
Eq. (\ref{StatiDicotomiciConAttrito}). 
In the lateral region, after the end
of the state that has pushed the trajectory outside the central region,
the slope assigned to the trajectory by the ensuing states have always
the same sign, either negative, for the top region, or positive, for
the bottom region. This is a consequence of the definition itself of
side regions and central stripe, the former being those characterized
by $|x|> W/\gamma$ and the latter referring $|x|< W/\gamma$. The final
effect is that the trajectories that have exited the central region,
are steadily brought back to it. We see that, on the contrary, in the
central region the trajectory fluctuates back and forth more or less as
it would do in the case of free diffusion. This is perhaps the
intuitive reason why the linear response theory of Section \ref{theo},
referring to a central region of extremely large size, makes the
system's statistics fall in the L\'evy basin of attraction as it does in
the free case \cite{mario}.

\subsection{Computer simulation}
\label{computer}
The computer simulations
are done generating the trajectories of
Eq.(\ref{xattimet}) with the assumption that the values
$W_{+}(x_{k}) = W - \gamma x_{k} $ and $W_{-}(x_{k}) = -W -\gamma x_{k}$
have the same probability. The distribution of sojourn times in one
of these two states is given
by the function $\psi(t)$ of Eq.(\ref{explicit}), which, in turn, is
realized by a suitable nonlinear transformation of a variable with a
uniform distribution in the interval $[0,1]$ obtained by a standard
routine for the generation of random numbers. It has to be pointed out
that the value of either $W_{+}(x_{k})$ or $W_{-}(x_{k})$ is determined
by the value that the variable $x$ has at the moment when a new state
is established and it is kept fixed until the end of the sojourn
in this state, namely its value is fixed in the interval $[t_{k},t_{k+1}]$.
Each trajectory is generated according to the integral of the equation
(\ref{withfeedback}) with the initial condition $x(0) =0$ and it
is observed until the time $T_{f} \equiv 20/\gamma >> T_R$. At this time
the value is recorded in the form of a histogram. The
distribution of the variable $x$ at this final time, assumed to
correspond to equilibrium, is obtained by generating a sufficiently large
number of trajectories and constructing a histogram.

\section{Theoretical prediction}
\label{theo}
We are now in a position to demonstrate that in the case of extremely weak
feedback the resulting equilibrium distribution is L\'{e}vy.
There are no compelling theoretical predictions in the case of strong feedback,
except the heuristic arguments used in section \ref{intro}.

According to the strategy adopted in Ref.\cite{bianucci}
the theoretical approach to friction implies
the treatment of the
response of the booster to an abruptly applied external perturbation. Using
the notation adopted in Ref. \cite{bianucci}
the solution to the equation of motion
	\begin{equation}
	\frac{\partial}{\partial t} \rho(\xi,t) = \Gamma(K)\rho(\xi,t),
	\label{abrupt}
	\end{equation}
where $\Gamma(K)$
denotes the operator driving the motion of the
probability distribution $\rho(\xi,t)$ in the presence of a
perturbation of intensity $K$. We assume that this operator can be
expressed as the sum of an unperturbed and a perturbed part,
	\begin{equation}
	\Gamma(K) = \Gamma_{0} + K \Gamma_{1},
	\label{perturbation}
	\end{equation}
where $\Gamma_{0}$ denotes the operator driving the unperturbed
motion of the variable $\xi$ and $K\Gamma_{1}$ denotes the
operator corresponding to the
perturbation. The intensity of this perturbation can be freely
changed by modifying the size of the parameter $K$. 
We apply the perturbation abruptly at time $t =0$, to
the booster, assumed to be in the equilibrium state, $\rho_{eq}$. 
The perturbation
strength is assumed to be so weak as to make it possible to solve
Eq.(\ref{abrupt}) at the first order perturbation,
	\begin{equation}
	\frac{\partial}{\partial t} \rho_{1}(\xi,t) = 
	\Gamma_{0}\rho_{1}(\xi,t) + K\Gamma_{1} \rho_{0},
	\label{firstorder}
	\end{equation}
where $\rho_{0}$ and $\rho_{1}$ denote the zero-th and the first-order 
distribution density, respectively.
We assume that the system before the abrupt perturbation is in 
thermodynamic equilibrium. This means
that we identify $\rho_{0}$ with $\rho_{eq}$,
which is assumed to fulfill the following equation:
	\begin{equation}
	\Gamma_{0}\rho_{eq} = 0
	\label{equilibrium}.
	\end{equation}
The formal  solution to Eq.(\ref{firstorder}) is
	\begin{eqnarray}
	\rho_{1}(\xi,t) & = & K \int_{0}^{t} dt' exp[\Gamma_{0}(t-t')] 
	\Gamma_{1} \rho_{0}(\xi,t')  = \nonumber \\
	& = & K \int_{0}^{t} dt' exp(\Gamma_{0}t') \Gamma_{1}
	\rho_{0}(\xi,t-t') = \nonumber \\
	& = & K \int_{0}^{t} dt' exp(\Gamma_{0}t') \Gamma_{1} \rho_{eq}.
	\label{solution}
	\end{eqnarray}
We assume that at equilibrium the mean value of $\xi$ vanishes. As a
consequence, Eq.(\ref{solution}) yields
	\begin{equation}
	\langle \xi(t) \rangle = K \int_{0}^{t}C(t')dt',
	\label{current}
	\end{equation}
where
	\begin{equation}
	C(t) \equiv \langle \xi exp(\Gamma_{0}t)\Gamma_{1} \rangle_{eq}.
	\label{keyfunction}
	\end{equation}
This concise expression means a trace over $\xi$ of the
distribution obtained by applying to $\rho_{eq}$ first the
perturbation operator $\Gamma_{1}$, then the unperturbed time
evolution operator $exp(\Gamma_{0}t)$ and finally the variable $\xi$
itself.

In the case of ordinary statistical mechanics one usually makes the
assumption that the variable $\xi$ has Gaussian statistics. This means a
continuous variable 
driven by both the unperturbed stochastic
environment and by the perturbation of intensity $K$.
To define the action of the perturbation on the time evolution of the
$\xi$-trajectory, let us switch off the influence of the stochastic
environment. In this case
the perturbation action would be expressed by
$d\xi/dt = K$. Moving from the Heisenberg-like to the
Schr\"{o}dinger-like
representation, we would obtain
$\frac{\partial}{\partial t} \rho(\xi,t) = K
\frac{\partial}{\partial \xi}\rho(\xi,t)$
which implies that the perturbation operator is
	\begin{equation}
	\Gamma_{1} = \frac{\partial}{\partial \xi}.
	\label{ordinary}
	\end{equation}
Note that in the Gaussian case $\rho_{eq}(\xi)$
is a Gaussian distribution proportional to
$exp(-\xi^{2}/(2\langle\xi^{2}\rangle_{eq}))$. Consequently
$C(t)$ of Eq.(\ref{keyfunction}) becomes
	\begin{equation}
	C(t) = \Phi_{\xi}(t).
	\label{identification}
	\end{equation}
Note that
	\begin{equation}
	d\langle x(t)\rangle/dt = \langle \xi(t)\rangle
	\label{current2}
	\end{equation}
and that, in the absence of perturbations,
with all the trajectories starting at $x=0$,
	\begin{equation}
	\langle x^{2}(t)\rangle_{0} = 2 \langle\xi^{2}\rangle_{eq} 
	\int_{0}^{t}dt' \int_{0}^{t'}dt'' \Phi_{\xi}(t'').
	\label{secondmoment}
	\end{equation}
We adopt the subscript $0$ to indicate that the time evolution of the
second moment takes place in the absence of perturbations.
Using Eq.(\ref{current2}), Eq.(\ref{secondmoment}), and
Eq.(\ref{current}) supplemented by
Eq.(\ref{identification}), and
setting equal to zero an arbitrary integration constant so as to ensure
a response proportional to the perturbation strength, we arrive at
	\begin{equation}
	\langle x(t)\rangle = K \langle x^{2}(t)\rangle_{0}/
       	(2\langle\xi^{2}\rangle_{eq}).
	\label{einstein}
	\end{equation}
This equation is a generalized Einstein
relation \cite{bouchaud}, more recently discussed by Barkai and
Fleurov \cite{barkai}.

It is evident that the ordinary condition, (\ref{einstein}), 
cannot apply in the case
here under study, corresponding to the conditions Eqs.
(\ref{timebreakdownofextensivity}) and (\ref{crucialpowerinterval}).
In fact, in this latter case the adoption of the ordinary linear response
relation would yield an infinite current.
 Earlier investigation \cite{trefan}
has established that the ordinary linear response theory,
breaks down,
and a new form of linear response to perturbation shows up, which,
however, cannot be expressed in terms of unperturbed dynamics. This
means that the response depends on the dynamical model adopted and
must be established through  numerical integration.

In conclusion, using the strategy of Ref.\cite{bianucci} we find that
in the presence of feedback the free diffusion process
	\begin{equation}
	dx/dt = \xi
	\label{freediffusion}
	\end{equation}
must be replaced with
	\begin{equation}
	dx/dt = \langle \xi(t)\rangle_{x} + \xi,
	\label{feedback}
	\end{equation}
where we have divided the current,
in accordance with linear response theory
into the sum of two terms. The zero-th order term
is given by the fluctuation $\xi$ assumed to be
in the same unperturbed condition as in Eq.(\ref{freediffusion}).
The first-order term is  $\langle \xi(t)_{x}\rangle$. Although the variable
$\xi$ sojourns for long times in one of the two-velocity states, we
assume the feedback to be so weak as to be compatible with $\xi$
making many jumps from one velocity state to the other while
the value of the variable $x$ remains essentially unchanged.
On the other hand, setting\cite{bianucci}
	\begin{equation}
	K = -\Delta^{2} x
	\label{assumption}
	\end{equation}
where $\Delta^2$ is a constant, we obtain
	\begin{equation}
	dx/dt = -\gamma x(t) + \xi(t),
	\label{langevin}
	\end{equation}
and
	\begin{equation}
	\gamma = \Delta^{2} \chi,
	\label{friction}
	\end{equation}
the susceptibility $\chi$ being defined by
	\begin{equation}
	\chi \equiv \int_{0}^{\infty}C(t)dt
	\label{susceptibility}.
	\end{equation}

We are now in a position to address the problem under discussion 
here via three approximation steps:

(i) We replace Eq.(\ref{langevin}) with
	\begin{equation}
	dx/dt = -\gamma x(t) + \eta(t),
	\label{langevin2}
	\end{equation}
where $\eta$ is a L\'{e}vy noise\cite{seshadri}. This means that the
variable $\eta$ in one single time step produces jumps proportional
to those produced by the variable $\xi$ sojourning for a time $t =
\eta/W$ in one of the two velocity states\cite{juri,mario}.

(ii) We make a numerical simulation of Eq. (\ref{langevin})

(iii) We make a completely dynamical treatment of the whole
fluctuation-dissipation process.

We shall refer to approximation (i) as the \emph{stochastic
approximation}. The advantage of this approximation is that
Eq.(\ref{langevin}) is made equivalent to an equation studied years ago by
West and Seshadri\cite{seshadri}.The stochastic force results in a
phase space operator equivalent to a fractional derivative
of order $\alpha$, with
	\begin{equation}
	\alpha = \beta + 1.
	\label{alpha}
	\end{equation}
Thus, it is possible to show\cite{seshadri} (see also
Refs.\cite{mauro,mario}
for a more recent discussion of the same problem) that the Fourier
transform of the distribution density $\sigma(x,t)$,
$\hat{\sigma}(k,t)$,reads:
	\begin{equation}
	\frac{\partial}{\partial t}\hat{\sigma}(k,t)
	= - \left(b|k|^\alpha +
	\gamma k \frac{\partial}{\partial k}\right)\hat{\sigma}(k,t),
	\label{fouriertransform}
	\end{equation}
which yields the following characteristic function
	\begin{equation}
	\mbox{ln} \hat{\sigma}(k,t) = -\frac{(1-e^{-\alpha\gamma t})}
	{\alpha\gamma} b |k|^\alpha
	\end{equation}
which for no dissipation has the familiar form 
	\begin{equation}
	\lim_{\gamma \rightarrow 0} \mbox{ln} \hat{\sigma}(k,t)
	= - t b |k|^\alpha .
	\end{equation}
However, asymptotically, in time we obtain the 
equilibrium condition
	\begin{equation}
	\hat{\sigma}(k,\infty) = exp\left( - \frac{b |k|^\alpha }{\alpha
	\gamma}\right)= e^{- b_\gamma |k|^\alpha }.
	\label{equilibrium2}
	\end{equation}
The second equality defines the equilibrium parameter $b_\gamma$
of the L\'evy characteristic function.
Approximation (ii) is referred to as the \emph{dissipative
L\'{e}vy walk},  and has already been
discussed in an earlier work\cite{juri,mario}, where it was shown that in
the limiting case of $\gamma \rightarrow 0$ leads to an equilibrium
distribution equivalent to that of Eq.(\ref{equilibrium2}).

\section{The numerical simulation}
\label{num-res}

An interesting result of the numerical simulation of the dynamical
model is that the equilibrium distribution is not uniquely
determined.
 The condition of weak feedback yields a distribution
which has properties different from those of the equilibrium
distribution stemming from the condition of strong feedback.


 The numerical simulation of the dynamical model can be done both
in the case of weak and strong feedback. Interesting new effects are
revealed by the simulation of the case of strong feedback.
However, these are left as a subject for future theoretical
discussions. Here we  illustrate only the simulation results concerning
the case of weak feedback. According to the program of Ref.\cite{bianucci}, we
have developed a theory resting on the linear response theory, although
in the non-conventional form of Section \ref{theo}, and consequently on the
assumption of a very weak feedback.  Therefore, the simulation results
here discussed refer to a case of feedback so weak as to ensure that
the requirements of Section \ref{theo} are fulfilled.

	\begin{figure}
	\begin{picture}(210,180)(0,0)
	\epsfysize=6truecm
	\epsfxsize=8.5truecm
	\put(-10,10){\epsfbox{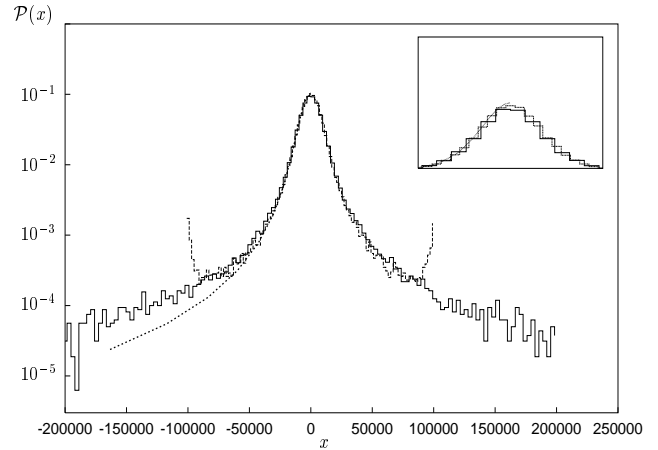}}
	\end{picture}
	\caption{
The equilibrium distribution in the L\'evy's basin of attraction. The full
line histograms refer to the numerical simulation with $\beta = 0.5$,
$\gamma = 10^{-5}$, $W = 1$, $\bar{T}=50$. We set the bin size 
equal to $2000$.  The dashed line histograms illustrate
the result of the numerical simulation of the dissipative L\'evy walk of
Eq. (\ref{langevin}). The prediction of the stochastic approximation, or,
equivalently, of the WS statistics, is denoted by means of the heavily
dashed line. To make the figure less heavy we plot only the left part
of the distribution predicted by the WS statistics. The WS equilibrium
is obtained evaluating the inverse Fourier transform of 
Eq.(\ref{equilibrium2}) with $b_\gamma = 417771$, $\alpha =1.5$ and 
$\gamma = 10^{-5}$. 
This is the value of $b_\gamma$ that
according to the theoretical prediction of Eq.(\ref{b_eq}) corresponds to the
parameters of the dynamical treatment.
The insert shows, for clarity, the enlarged portion 
of the figure corresponding to the 
$x$-axis interval $[-20000,20000]$. The enlargement 
of the ordinates is done after conversion to a linear scale.}
	\label{LevyDinamico}
	\end{figure}
Fig. \ref{LevyDinamico} refers to the L\'evy basin of attraction. 
We notice that the dissipative L\'evy walk results in pronounced peaks. 
These peaks are produced by the fact that the trajectories cannot 
overcross the values $x = \pm W/\lambda$. These peaks signal the 
region within which a good agreement among the fully dynamic treatment, the
dissipative L\'evy walk and the stochastic approximation  is expected. 
Note that, as remarked in Section \ref{theo},  the stochastic approximation
is equivalent  to the equilibrium of the theory of West and Seshadri (WS)
\cite{seshadri}, where the parameter $b_\gamma$ of Eq. (\ref{equilibrium2})
is calculated with the following formula \cite{mario}:
	\begin{equation}
	b_\gamma =\frac{W(\beta \bar{T} W)^\beta}{\gamma (\beta+1)} 
	\sin\left(\frac{\pi}{2}\beta \right) \Gamma(1-\beta),
	\label{b_eq}
	\end{equation}
where $\bar{T}$ and $\beta$ are defined in Eq.(\ref{correlationfunction}) 
and $\Gamma(\cdot)$ is the Euler gamma function.
We see that in the region enclosed by the peaks a very good 
accordance among these three distinct approaches is found indeed.

	\begin{figure}
	\begin{picture}(210,180)(0,0)
	\epsfysize=6truecm
	\epsfxsize=8.5truecm
	\put(-10,10){\epsfbox{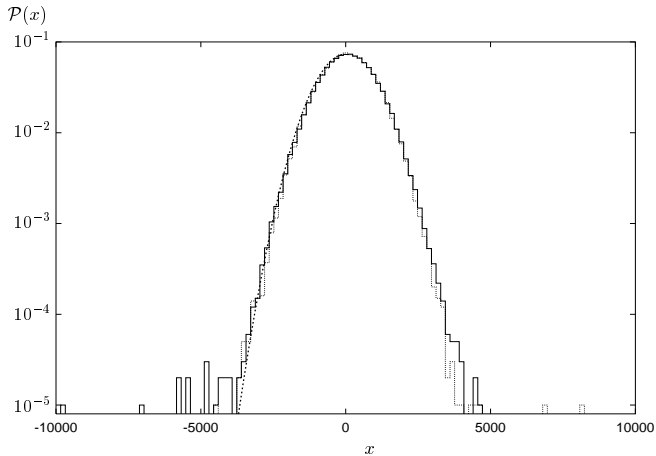}}
	\end{picture}
	\caption{
The equilibrium distribution in the Gaussian basin of
attraction. The full line histograms refer to the numerical simulation
with $\beta = 3$, $\gamma = 10^{-4}$, $W = 1$, $\bar{T} = 50$.
We set the bin size equal to $120$.
The dashed line histograms refer to 
the numerical solution of the Gaussian counterpart
of Eq. (\ref{langevin}). The heavily dashed line is the theoretical 
prediction of the Ornstein-Uhlenbeck process with
$\sigma^2=7.5\cdot 10^5$ calculated from Eq. (\ref{sigma_eq}).
To make the figure less 
heavy we plot only the left portion of this theoretical prediction.}
	\label{GaussianaDinamica}
	\end{figure}
    Fig. \ref{GaussianaDinamica} illustrates the situation in the Gauss 
basin of attraction. In this case the side peaks of the dissipative walk
disappear. The counterpart of the WS statistics here is the 
Ornstein-Uhlenbeck statistics \cite{seshadri}, already used in 
an earlier publication \cite{mario} to predict the form of equilibrium 
distribution in the Gauss basin of attraction. 
We use the same prediction where the variance $\sigma^2$ is:
	\begin{equation}
	\sigma^2 = \frac{\beta W^2 \bar{T}}{\gamma (\beta -1)}
	\label{sigma_eq}
	\end{equation}
and we find that the agreement with the other two treatments 
is again very good. 

       We are convinced that the simulation results of 
Fig.\ref{LevyDinamico} are a reliable numerical evidence of the fact that 
the dynamic approach to non-canonical equilibrium yields the analytical
form proposed 18 years ago by West and Seshadri \cite{seshadri}, 
here referred to as West-Seshadri (WS) statistics. However, 
Tsallis statistics and WS statistics have in
common the analytical shape of the tails, which is an inverse power
law. This might suggest that the accordance between our simulation
results and Tsallis statistics is as satisfactory as, or more
satisfactory than, the accordance with the WS statistics. We now show
that it is not so, and that the accordance of our results
with Tsallis statistics is much less satisfactory than with the  WS
statistics. According to the spirit of our dynamic approach, the
comparison should be done with the Tsallis form of equilibrium
corresponding to a constraint on the first moment of $|x|$. This would
produce a more significant departure from the WS statistics, and from
our simulation results, the WS statistics being, in fact, a L\'evy form of
equilibrium statistics. The reader can easily convince him/herself
about this property observing Fig \ref{ConfrontoLevyTsallis}.  
Thus, we decided to discuss the comparison between our simulation results 
and Tsallis statistics adopting  a condition more favorable to the  Tsallis
statistics, namely, the analytical proposal of Eq.(\ref{new}).

	\begin{figure}
	\begin{picture}(210,180)(0,0)
	\epsfysize=6truecm
	\epsfxsize=8.5truecm
	\put(0,10){\epsfbox{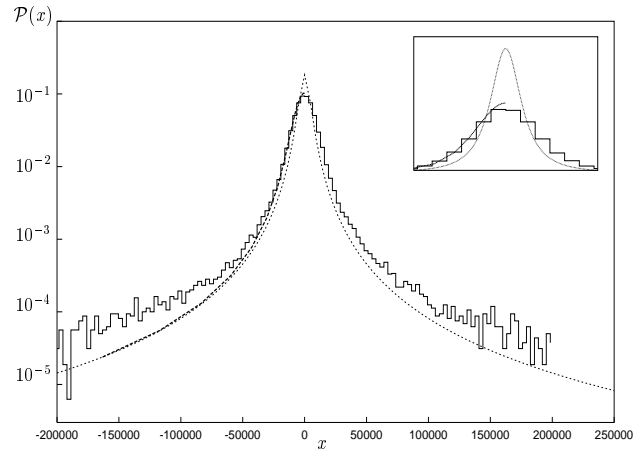}}
	\end{picture}
	\caption{
Tsallis and WS statistics versus the numerical results. The WS
equilibrium is denoted by the heavily dashed line and, as in 
Fig. \ref{LevyDinamico},
only the left part of it is illustrated. It corresponds to the inverse
Fourier transform of the distribution of Eq. (\ref{equilibrium2}) 
with $b_\gamma = 417771$, $\alpha = 1.5 $ and $\gamma = 10^{-5}$. 
The parameter $b_\gamma$ was found as in Fig. \ref{LevyDinamico}.
The  Tsallis equilibrium, corresponding to the
proposal of  Eq. (\ref{new}), is illustrated by the dotted line. The dotted
line is plotted so as to coincide with the heavily dashed line in the
region of large distances ($\tilde{b}=6\cdot 10^{-8}$, $q=1.8$).
The insert shows, for clarity, the enlarged portion 
of the figure corresponding to the 
$x$-axis interval $[-20000,20000]$. The enlargement 
of the ordinates is done after conversion to a linear scale.}
	\label{accordo1}
	\end{figure}
\begin{figure}
	\begin{picture}(210,180)(0,0)
	\epsfysize=6truecm
	\epsfxsize=8.5truecm
	\put(0,10){\epsfbox{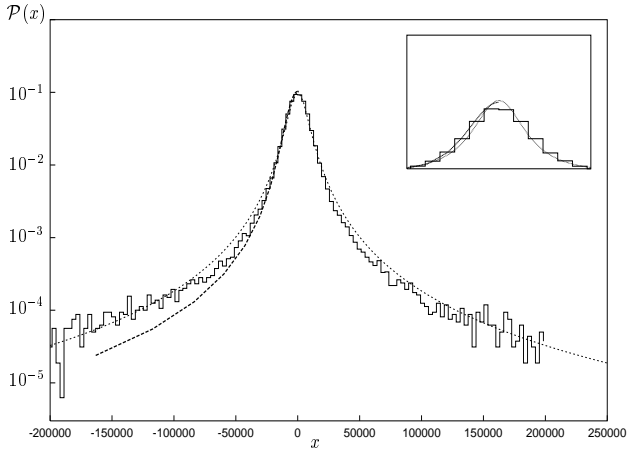}}
	\end{picture}
	\caption{
Tsallis and WS statistics versus the numerical results. The WS
equilibrium is denoted by the heavily dashed line and, as in 
Fig. \ref{LevyDinamico}, only the left part of it is illustrated. 
It corresponds to the inverse
Fourier transform of the distribution of Eq. (\ref{equilibrium2})  
with $b_\gamma = 417771$, $\alpha  = 1.5 $ and $\gamma = 10^{-5}$. 
The  Tsallis equilibrium, corresponding to the
proposal of  Eq. (\ref{new}), is illustrated by the dotted line
($\tilde{b}=2\cdot 10^{-8}$, $q=1.8$). 
The dotted line is plotted so as to get the best 
fitting with the heavily dashed line in the central region.  
The insert shows, for clarity, the enlarged portion 
of the figure corresponding to the 
$x$-axis interval $[-20000,20000]$. The enlargement 
of the ordinates is done after conversion to a linear scale.}
	\label{accordo2}
	\end{figure}

  In Fig.\ref{accordo1} we establish an exact accordance between Tsallis and
WS statistics in the region of large distances. We see that the WS
statistical prescription  yields a very satisfactory accordance with
the results of simulation also in the central region, while the Tsallis
non-canonical equilibrium does not. In Fig.\ref{accordo2} we organize the
comparison in such a way as produce the best fitting between Tsallis
statistics and our simulation in the central part of the
equilibrium distribution. We see that this has the effect of making the
discrepancy between Tsallis statistics and WS statistics much worse in
the region of large distances. 
We see that our simulation results are much closer to the WS equilibrium
than to the Tsallis equilibrium. We have to point out, furthermore,
that the region of large distances of Fig. \ref{accordo2}, where the 
Tsallis form of non-canonical equilibrium apparently yields a better 
agreement with our numerical results, is probably a non-stationary 
region, which depends on the observation time. We do not have yet any 
theory concerning this region but the result of the numerical observation. 
The numerical simulation shows that this region triggers non-stationary 
effects, due to the fact that the trajectories reaching large distances 
from the central diffusion region, illustrated in 
Fig. \ref{CampioneProcessoConRetroazione}, tend to escape
forever. This escape process gives rise, in a time scale that becomes
infinitely large upon decrease of the feedback strength, to a diffusion
process, namely to a form of non-equilibrium distribution.

We are convinced that the WS is closer to the results of numerical 
simulation than the Tsallis non-canonical equilibrium. We see, however, 
that in the case of Fig. \ref{accordo2} the departure of the Tsallis 
prediction from 
the results of the numerical simulation is not so marked as to rule out 
this theoretical proposal. In conclusion, if we use only fitting 
arguments,  we can select the constraint on the second moment rather 
than that on the first, as the adoption of theoretical arguments would 
suggest us to do. Eventually, the situation would appear as favorable 
to Tsallis statistics as illustrated in Fig. \ref{accordo2}.


\section{Concluding remarks}
\label{concl}

The present paper is not just about obtaining the best fit to a 
numerical simulation. We aim at deriving 
equilibrium from dynamics with no use of thermodynamic arguments. 
The result of our investigation is that equilibrium is dictated by 
the WS statistics rather than by the the Tsallis statistics. This 
yields a satisfactory agreement with the results of numerical 
simulation with no fitting parameters. The advocates of Tsallis 
statistics might reach a result as satisfactory as that provided 
by the dynamic theory (even if we think that actually it is less 
satisfactory) only using fitting parameters. This paper results 
in two important facts. The first is a numerical simulation yielding 
a non-canonical equilibrium. The second is a theory to account 
for this non-canonical equilibrium. The theoretical arguments lead 
us to conclude that the non-canonical equilibrium must correspond 
to the WS statistics. 

If only fitting arguments were to be used, it would be more difficult 
to make a choice between the Tsallis and the WS statistics. 
However, if we invoke 
also theoretical arguments, we find that no room is left for the 
Tsallis statistics in the case of the dynamic model studied in the 
present paper. At least within the range of this dynamic model, we show 
that the non-canonical equilibrium is possible in nature, but it must 
correspond to the theoretical proposal of West and Seshadri 
\cite{seshadri} rather 
than to the predictions of non-extensive thermodynamics 
\cite{tsallis1}. 
It is 
interesting to remark that in the limiting case of very weak feedback 
our model becomes identical to the L\'evy flights subject to the Hookean 
force of the recent work of Jespersen, Metzler and Fogedby \cite{jespe}, and 
that both models are equivalent to that originally studied by West and 
Seshadri \cite{seshadri}. 
Also these authors \cite{jespe} found the Tsallis statistics to 
be incompatible with non-Gibbsian nature of the corresponding 
stationary solution.

In a lucid discussion Lebowitz \cite{lebowitz} has recently restated
the point of view of Boltzmann, establishing the microscopic origin of
irreversible macroscopic behavior.
In his view the adoption of the laws of large numbers is essential, and
the role of deterministic chaos becomes important only if it applies
to a macroscopic  number of non-interacting particles.
According to Lebowitz, mixing and ergodicity are notions that are
``unnecessary, misguided and misleading.'' In other words, this opinion
reflects the conviction, mirrored by the handbooks  of statistical
mechanics, that
the unification of mechanics and
thermodynamics rests on the condition $N\rightarrow \infty$, where
$N$ denotes the number of degrees of freedom of the system under
study.
These statements do not, however, conflict with the results of
Ref.\cite{bianucci} reported in Section \ref{dyntotherm}. 
In fact, the signature
of the dynamical origin of thermodynamics, as expressed by
Eq.(\ref{bianucci}), is lost in the limiting case of 
the thermodynamic limit $N \rightarrow\infty$, thereby 
making the controversy between the advocates of
mixing and the advocates of $N = \infty$ difficult,
if not impossible, to substantiate
with experimental arguments.
This means, in other words, that the canonical equilibrium distribution
can be derived using simple arguments, based only on
probabilistic concepts and the law of large numbers, or, if we wish,
also on the dynamical arguments of Ref.\cite{bianucci}. If the criterion
of simplicity is adopted, one might be tempted to choose the former
approach which leads to the wanted result with little or no effort, while the
second approach yields the same conclusion after many
complicated calculations based on assumed dynamical properties, that
only in a few cases have been rigorously proved.

        The present paper shows how to extend  to the case of boosters with no
finite time scale the program of Ref. \cite{bianucci}. In this new case the
dynamical approach to equilibrium yields a non-canonical equilibrium
which is that advocated many years ago by West and Seshadri \cite{seshadri}. 
As pointed out by the theoretical discussion of Section \ref{theo}, it must be
remarked that Eq. (\ref{langevin2}), yielding the form of equilibrium 
of West and Seshadri, is the consequence of a form of linear response, 
departing from the traditional wisdom behind the generalized Einstein 
relation \cite{bouchaud}. 
It is remarkable that the only plausible form of linear response,
resting on dynamics, yields the same prescription as that suggested by
a phenomenological approach. Thus, while we do agree with Rajagopal and
Abe about the fact that the canonical  equilibrium is not the only
acceptable form of equilibrium, we depart from them on the specific
form that this equilibrium will take, since, as we have seen in Section
\ref{non-ext}, the L\'evy statistics must not be confused with the generalized
canonical distribution of Tsallis.



\begin{references}


\bibitem{tsallis1} C. Tsallis, J. Stat. Phys. {\bf 52}, 479 (1988).


\bibitem{tsallis2} C. Tsallis, R.S. Mendes, and A.R. Plastino,
Physica A {\bf 261}, 534 (1998).

\bibitem{brazil}
Nonextensive statistical mechanics aims at offering a theoretical
framework for systems with long-range
interaction, long-range memory, or fractal structure. For its
application, see, for instance, Br. J. Phys. {\bf 29}, special issue
(1999). A comprehensive list of reference is currently available at
http://tsallis.cat.cbpf.br/biblio.htm

\bibitem{walton} D.B. Walton and J. Rafelski, Phys. Rev. Lett. {\bf
84}, 31 (2000).

\bibitem{wilk} G. Wilk and Z. W\l odarczyk, hep-ph/9908459

\bibitem{abe}  S. Abe, A. K. Rajagopal, Phys. Lett. A {\bf 272}, 341 (2000).

\bibitem{katz} A. Katz, \emph{Principles of Statistical Mechanics: the
information theory approach}, Freeman,  San Francisco (1967).


\bibitem{thirdrule} C.Tsallis, R.S. Mendes and A.R. Plastino, Physica A
{\bf 261}, 534 (1998).

\bibitem{note1} The rule prescribed by Ref.\cite{thirdrule} refers to
the case where $x$ is the energy of the system. However, the approach
adopted by these author rest on information theory \cite{katz}, and for
this
reason we can apply it to any ``stochastic'' variable $x$, whatever
is physical meaning.

\bibitem{anna} M. Buiatti, P. Grigolini, A. Montagnini,
   Phys. Rev. Lett. {\bf 82}, 3383 (1999).

\bibitem{paolonote} The dynamic approach to L\'evy statistics \cite{anna} 
is based on the the
equation of motion $dx/dt = \xi(t)$, where $\xi(t)$ is a stochastic            
variable with only two values, $W$ and $-W$. The entropic  argument is
used to derive the waiting time distribution $\psi(t)$ with a constraint
on the first moment of $t>0$. Thus, the function $\Pi(x)$ is related to
$\psi(t)$ by the key relation: $\Pi(x) = (1/2)\psi(|x|/W)/W$. 
Therefore,  it is now evident that the dynamic approach to L\'evy 
statistics implies a constraint on $|x|$.

\bibitem{MS84} E.W. Montroll and M.F. Shlesinger, in {\em Nonequilibrium
Phenomena II, ``From Stochastics to Hydrodynamics''}, edited by J.L.
Lebowitz and E.W. Montroll (North-Holland, Amsterdam, 1984).

\bibitem{bruce} E.W. Montroll and B.J. West, in {\em Fluctuation
Phenomena}, 2nd ed., edited E.W. Montroll and J.L.
Lebowitz, Studies in Statistical Mechanics
Vol. 7  (North-Holland, Amsterdam, 1987).

\bibitem{gnekol} B.V. Gnedenko and A.N. Kolmogorov, 
	\emph{Limit Distributions for Sums of Independent
	Random Variables}, Addison-Wesley Publishing Company, Cambridge.

\bibitem{bianucci} M. Bianucci, R. Mannella, B.J. West, P. Grigolini,
Phys. Rev. E {\bf 51}, 3002 (1995).

\bibitem{ford} G.W. Ford, M. Kac, and P. Mazur, J. Math. Phys. {\bf 6}, 
504 (1965).

\bibitem{ullersma} P. Ullersma, Physica (Utrecth) {\bf 32}, 27 (1966);
{\bf 32}, 56 (1966); {\bf 32}, 74 (1966).

\bibitem{philipson} P.E. Philipson, J. Math. Phys. {\bf 15}, 2127
(1987).

\bibitem{vitali} D. Vitali and P. Grigolini, Phys. Rev. A {\bf 39},
1486 (1989).

\bibitem{lindenberg} K. Lindenberg and B. J. West, \emph{The
Nonequilibrium Statistical Mechanics of Open and Closed Systems}, VCH
Publishers, New York (1990).

\bibitem{lindenberg2} See Lindenberg and West \cite{lindenberg}
for a complete review.

\bibitem{huang} K. Huang, \emph{Statistical Mechanics}, John Wiley,
New York (1987).

\bibitem{fermi} E. Fermi, J. Pasta and S. Ulam, in \emph{Collected
Works of Enrico Fermi}, University of Chicago Press, vol. II, p. 978
(1965).

\bibitem{allegro} P. Allegrini, P. Grigolini, and B.J. West, Phys. Rev. E
{\bf 54}, 4760 (1996).

\bibitem{lebowitz} J.L. Lebowitz, Physica A {\bf 263}, 516 (1999).

\bibitem{gross} D.H.E. Gross, E. Votyakov, cond.mat/9911257.


\bibitem{zaslavsky}  G. M. Zaslavsky,  {\em Physics of Chaos in Hamiltonian
Systemson the Foundations of Statistical Physics},Imperial College Press,
London (1998).

\bibitem{geisel} T. Geisel and S. Thomae, Phys. Rev. Lett. {\bf 52}, 1936
(1984).



\bibitem{juri} M. Annunziato, P. Grigolini, J. Riccardi,
	 Phys. Rev. E {\bf 61} (2000) 4801.

\bibitem{mario} M. Annunziato, P. Grigolini, Phys. Lett. A
	{\bf 269} (2000) 31.


\bibitem{mauro}M. Bologna, P. Grigolini, and J. Riccardi,
   Phys.Rev. E {\bf 82} , 6432 (1999).

   \bibitem{gaspard} P. Gaspard, X.-J. Wang, Proc. Nat. Acad.Sci.USA
   {\bf 85}, 4591 (1988).

   \bibitem{grigo} M. Ferrario, P. Grigolini, A. Tani, R. Vallauri,
   B. Zambon,Adv. Chem. Phys. {\bf 62}, 225 (1985).

\bibitem{thechoiceofmario} T.Geisel, J. Helstab and H.Thomas, Z. Phys.
B {\bf 55}, 165 (1984).

\bibitem{tesimario} M. Annunziato, Ph.D. Thesis, {\it unpublished}.

\bibitem{bouchaud} J.P. Bouchaud and A. Georges, Phys. Rep. {\bf
195}, 127 (1990).

\bibitem{barkai} E.Barkai and V.N. Fleurov, Phys. Rev. E
{\bf 58}, 1296 (1998).

\bibitem{trefan} G. Trefan, E. Floriani, B.J. West, and P. Grigolini,
Phys. Rev. E {\bf 50}, 2564 (1994).

\bibitem{seshadri} B.J. West and V. Seshadri, Physica A {\bf 113},
203 (1982).

\bibitem{jespe} S. Jespersen, R. Metzler, H.C. Fogedby, Phys. Rev. E
{\bf 59}, 2736 (1999).


\end{references}
\end{document}